\documentclass[12pt,preprint]{aastex}
\usepackage{bbm}
\usepackage{mathrsfs}
\usepackage{amssymb,amsmath,float}
\usepackage{rotating}
\usepackage{color}

\newcommand\al{\alpha}

\newcommand\de{\delta}

\newcommand\et{\eta}

\newcommand\lam{\lambda}
\newcommand\rh{\rho}
\newcommand\si{\sigma}

\newcommand\vp{\varphi}

\newcommand\om{\omega}

\newcommand\De{\Delta}

\newcommand\pa{\partial}
\newcommand\bna{\bold{\nabla}}
\newcommand\<{\langle}
\renewcommand\>{\rangle}

\newcommand\ie{\emph{i.e.}}
\newcommand\eg{\emph{e.g.}}

\newcommand\beq{\begin{equation}}
\newcommand\eeq{\end{equation}}
\newcommand\bea{\begin{eqnarray}}
\newcommand\eea{\end{eqnarray}}
\newcommand\bal{\begin{align}}
\newcommand\eal{\end{align}}

\newcommand\fr{\frac}

\newcommand\ap{\approx}
\newcommand\cd{\cdot}

\newcommand\bj{\bold{j}}

\newcommand\br{\bold{r}}

\newcommand\bA{\bold{A}}
\newcommand\bB{\bold{B}}

\newcommand\bE{\bold{E}}

\renewcommand\bal{\mbox{\boldmath$\alpha$}}

\newcommand\cA{\mathcal{A}}

\begin{document}

\title{Does light from steady sources bear any observable imprint of the dispersive intergalactic medium?}

%\author{Richard Lieu$^1$}
\author{Richard Lieu$^1$, Lingze Duan$^1$}

\affil{$^1$Department of Physics, University of Alabama,
Huntsville, AL 35899\\}

\begin{abstract}
There has recently been some interest in the prospect of detecting ionized intergalactic baryons by examining the properties of incoherent light from background cosmological sources, namely quasars.  Although the paper by \cite{lieu13} proposed a way forward, it was refuted by the later theoretical work of \cite{hir14} and observational study of \cite{hal16}.  In this paper we investigated in detail the manner in which incoherent radiation passes through a dispersive medium both from the frameworks of classical and quantum electrodynamics, which led us to conclude that the premise of \cite{lieu13} would only work if the pulses involved are genuinely classical ones involving many photons per pulse, but unfortunately each photon must not be treated as a pulse that is susceptible to dispersive broadening.  We are nevertheless able to change the tone of the paper at this juncture, by pointing out that because current technology allows one to measure the phase of individual modes of radio waves from a distant source, the most reliable way of obtaining irrefutable evidence of dispersion, namely via the detection of its unique signature of a quadratic spectral phase, may well be already accessible.  We demonstrate how this technique is only applied to measure the column density of the ionized intergalactic medium.

%For now, this technique is only applied to measure the column density of the ionized interstellar medium, because expansion and redshift complicates the calculation on larger scales, which has to be tackled in a separate paper.

\end{abstract}

\section*{Introduction}

Recently there has been a discourse in the literature on the measurability of the dispersion of light from steady extragalactic sources, mostly quasars, by the intervening ionized intergalactic medium, \cite{lie13}, \cite{lieu13}, and \cite{hir14}.  The topic is of interest to cosmology because if one could infer the line-of-sight plasma column density between the source and the observer, the sheer number and uniformity in the distribution of quasars would enable one to reconstruct the volume density of intergalactic plasma in various directions and redshifts, which is an invaluable resource for addressing one of the great enigmas of cosmology, namely the question of the whereabouts of the baryons in the near Universe, \cite{cen99}, \cite{dav01}.

The issue at stake here is as follows.   On one hand, lines-of-sight dispersion are undoubtedly measurable effects for unsteady extragalactic sources with clear time signatures like fast radio bursts (\cite{tho13}) which readily afford one with column density data for the relevant directions and redshifts, the directions to such sources are few and far between compared to quasars, and the source redshifts are usually unknown.  On the other hand, although quasars are much more numerous, they are not sufficiently unsteady on the short timescales where the imprints of dispersion are detectable using conventional and well established methods, or for that matter any method.  Indeed, while \cite{lie13} and \cite{lieu13} suggested possible ways forward, \cite{lieu13} already pointed out why \cite{lie13} cannot actually work and provided an improved way to reinstate it, \cite{hir14} argued that the assumptions underlying \cite{lieu13} itself are also flawed, and that in practice the dispersion of stationary light leaves behind no observable imprint in the context of the two 2013 papers, not even in terms of the microscopic statistical fluctuations of the light as compared to light from the same steady sources that did not pass through a dispersive medium.  The dispute between \cite{lieu13} and \cite{hir14} was eventually settled by an analysis of real observational data in the radio by \cite{hal16}, which led to the conclusion that either the intergalactic medium has far fewer baryons than the prediction of cosmological models, or \cite{hir14} was right in asserting that the ideas of \cite{lieu13} could not work.  The latter is much more likely to be the truth, as we shall demonstrate.

In this paper we provide a detailed treatment of the effect of an intervening plasma column on stationary chaotic light, \ie~incoherent radiation from a steady source with no preferred time stamp, to explain in both classical and quantum mechanical terms why \cite{lieu13} must fail, unless the source is at least partially coherent on certain scales (some quasars could exhibit this characteristic).  However, we defend \cite{lieu13} somewhat by pointing out an ambiguity in the full quantum approach to the problem that does not allow one to tell {\it a priori} whether \cite{lieu13} or \cite{hir14} is the correct model of the real situation, and only an observational effort such as \cite{hal16} can offer a resolution (although there are also other experiments serving the same purpose of discernment).  We then proceed to discuss yet another {\it new} endeavor to distinguish between stationary chaotic light that has and has not passed through a dispersive medium, using the framework that generalizes the \cite{hir14} interpretation, and argue that current technology in radio astronomy can put the idea into action.

The plan of the paper is that we shall begin by reviewing the formalism for describing an electromagnetic wave.  This is followed by a statement of the problem of photon noise, as we look for the simplest formulation that contains all the essentials.  After that, we shall tackle the phenomenon of dispersion semi-classically, before going on to a full quantum treatment and discuss the ambiguity that awaits, and how it may be resolved.  Lastly, we address the ramifications, and how to rectify them with a new observational strategy.

\section{Formalism}

As preamble, it is useful to recapitulate the standard quantum and classical frameworks, so that the notation and ansatz are both clear.  We shall henceforth work in units  $\hbar=c=1$.

\subsection{The intensity}

Classically an electromagnetic field in free space (where the free charge density is $\rh=0$) may be represented by a vector potential $\bA(t,\br)$ satisfying the wave equation and the transversality condition $\bna\cd\bA=0$.  The electric and magnetic fields are given by
 \beq \bE=-\fr{\pa\bA}{\pa t}, \qquad \bB=\bna\wedge\bA. \eeq
In much of what follows we will consider a one-dimensional problem, where the field is described by a potential with a single component, say $A_x(t,z)=\cA(t-z)$.  Then the non-vanishing field components are $E_x=B_y=-\cA'(t-z)$, where the prime denotes the derivative with respect to the argument of the function.  The Poynting vector lies in the $z$ direction and its magnitude (the intensity) is
 \beq I= |\bE\wedge\bB|=E_x B_y=[\cA'(t-z)]^2. \label{intensity}\eeq

In the quantum treatment, we have a vector potential operator $\hat\bA(t,\br)$ which again satisfies the wave equation and transversality condition.  In the one-dimensional case, it may be written in terms of creation and annihilation operators in the form
 \beq \hat A_x(t,z) = \hat A^{(+)}_x(t,z) + \hat A^{(-)}_x(t,z)
 = \int_0^\infty \fr{d\om}{\sqrt{4\pi\om S}}\, [\hat a(\om)e^{-i\om(t-z)}+\hat a^*(\om)e^{i\om(t-z)}], \label{A-A+} \eeq
where $S$ is the cross-sectional area of the beam.  ($S$ has to appear for correct normalization; in effect we are assuming that the wave is confined within a tube of cross-section $S$, and restricting our consideration to the modes with no transverse momentum).  Here the creation and annihilation operators satisfy the commutation relations
 \beq [\hat a(\om),\hat a^*(\om')] = \de(\om-\om'). \eeq
It is easy to verify that $\hat\bE$ and $\hat\bA$ are conjugate variables, namely \beq [\hat E_x(t-z), \hat A_x(t-z')]=\fr{i}{S}\de(z-z'). \eeq

The energy flux is again given by (\ref{intensity}), except that we must pay attention to the operator ordering problem.  Under most scenarios, we are interested in measurements over intervals long compared to the period of the wave, \ie~we may use a cycle-averaged intensity.  In that case, we only need to consider products of a positive and a negative frequency function (not two positives or two negatives, which would be very rapidly oscillating).  In a quantum formulation, we have to ensure that negative-frequency parts appear to the left of positive-frequency ones, so
 \bea \hat I(t) \!\!\!&=&\!\!\!  \hat E_x^{(-)}(t)\hat B_y^{(+)}(t)+\hat B_y^{(-)}(t)\hat E_x^{(+)}(t) \notag\\
 \!\!\!&=&\!\!\! 2\hat A_x^{(-)}{}'(t)\hat A_x^{(+)}{}'(t) \notag\\
 \!\!\!&=&\!\!\!  \fr{1}{2\pi S} \int d\om\,d\om'\,\sqrt{\om\om'}\hat a^*(\om)\hat a(\om')
 e^{i(\om-\om')t}. \label{Iquant} \eea

\subsection{Photon number flux}

Often, we are interested not in measurements of the energy flux, but in photon-number counts, so it is also useful to define a photon-number flux $\hat M(t)$, the rate at which photons are crossing the area $S$ at time $t$.  In the case where we are dealing a one-dimensional beam with a narrow band of frequencies, there is a useful way to do this, described \eg~by section 6.2 of \cite{lou00}.  Strictly speaking the integrations over $\om$ and $\om'$ in (\ref{Iquant}) run only from $0$ to $\infty$.  However, there is no reason why we should not use $k$ (= $k_z$) rather than $\om$ as the independent variable, and we could then include negative as well as positive values, representing waves travelling in the opposite direction.  Obviously for narrow-band beams we shall assume that those modes with negative $k$ are unoccupied.  Equivalently, we can simply extend the integration over $\om$ to run over the whole real axis.  Then we may define annihilation and creation operators $\hat a(t)$ and $\hat a^*(t)$ for photons at time $t$ by
 \beq \hat a(t) = \int_{-\infty}^\infty \fr{d\om}{\sqrt{2\pi}} \hat a(\om) e^{-i\om t}, \qquad
 \hat a^*(t) = \int_{-\infty}^\infty \fr{d\om}{\sqrt{2\pi}} \hat a^*(\om) e^{i\om t}. \label{aa*} \eeq
They clearly satisfy the commutation relation
 \beq [\hat a(t), \hat a^*(t')] = \de(t-t'). \label{atcomm}\eeq

The rate at which photons cross the area $S$, located at $z$, at time $t$ is then given by the operator $\hat M(t-z)$ where
 \beq \hat M(t) = \hat a^*(t) \hat a(t), \label{M} \eeq
so during a time interval $[t_0,t_0+\De t]$ the number of photons crossing the area is
 \beq \int_{t_0}^{t_0+\De t} dt\, \hat M(t). \eeq
Note that if we integrate over all time, we obtain
 \beq \int_{-\infty}^{\infty} dt\,\hat M(t) = \int_0^\infty d\om\,\hat a^*(\om)\hat a(\om), \eeq
i.e., the total photon number in the beam, as it should be.

It is also possible to express $\hat M (t)$ in terms of the potentials.  From (\ref{A-A+}) one finds \beq 2\hat A^{(-)} (t) \hat A^{(+)} (t) = \fr{1}{2\pi S} \int \fr{d\om\,d\om'}{\sqrt{\om\om'}}\hat a^*(\om)\hat a(\om') e^{i(\om-\om')t}.  \eeq  If the radiation has a dominant frequency $\om_0$, due to a spectral line or a narrow band filter, then to a good approximation $\sqrt{\om\om'} \ap \om_0$ and, using (\ref{aa*}), one obtains \beq \hat M(t) = 2S\om_0 \hat A^{(-)} (t) \hat A^{(+)} (t). \label{MAA} \eeq

\subsection{Coherent states}

In relating the classical and quantum calculations, coherent states play an important role.  The essential point is this.  For every classical solution of Maxwell's equations described by a vector potential $\cA(t,\br)$ there exists a corresponding coherent state $|\cA\>$, unique up to a phase factor, which satisfies the eigenvalue equation
 \beq \hat{\bA}^{(-)}(t,\br)|\cA\> = \cA^{(-)}(t,\br)|\cA\>, \eeq
and correspondingly
 \beq \<\cA|\hat{\bA}^{(+)}(t,\br) = \cA^{(+)}(t,\br)\<\cA|, \eeq
Thus the expectation value of the field operator in this state is precisely this solution:
 \beq \<\cA|\hat{\bA}(t,\br)|\cA\> = \cA(t,\br).\eeq
Owing to the preceding equations, this last result extends to any \emph{normally ordered} product of field operators.

If $\cA$ is the classical solution generated by a particular current distribution $\bj(t,\br)$, then in the quantum theory the application of this classical current distribution to the vacuum state will generate the corresponding coherent state, $|\cA\>$.

In the case of a one-dimensional beam, we can write
 \beq \cA^{(+)}(t,z)=\int_0^\infty \fr{d\om}{\sqrt{4\pi\om S}}\al(\om)e^{-i\om(t-z)}, \eeq
with a corresponding expression for $\cA^{(-)}(t,z)$ in terms of $\al^*(\om)$.  Then the coherent states satisfy
 \beq \hat a(\om)|\cA\>=\al(\om)|\cA\>, \qquad \<\cA|\hat a^*(\om)=\al^*(\om)\<\cA|. \label{ketbra} \eeq

\section{The problem}

The question to be addressed is whether by observing light from continuous sources propagating to us through a dispersive medium we can detect the effects of dispersion --- as we certainly can with pulsed sources.  For the moment, let us deal with the problem semiclassically.

There are three parts to the set-up, the source, the detector, and the propagation of the wave through the intergalactic medium.

\subsection{The source}

We may think of the distant source as a very bright disc (or a hemisphere or any other small shape) comprising many independent sources, such as electrons or atoms, centered for convenience on the origin.  Importantly, the radius of the disc, though macroscopic, is very small compared to its distance from the observer, which will allow us later on to use a one-dimensional plane wave representation.  Thinking classically for the moment, by an `independent source' one means a source that emits coherent radiation over some (generally short) period of time.  For example, if we are talking about an excited atom in a gas, it will count as a single independent source between one collision and the next; after that, because the phase changes discontinuously and randomly at each collision, it will count as a different source.

Classically, each source can be modeled as a current distribution, say $\bj_s(t,\br)$ for the source number $s$.  Here each $\bj_s$ will be non-zero only in some small region of space and interval of time, and will typically be oscillating with some characteristic frequency, modulated by a slowly varying amplitude function that defines one pulse of emission from the source.  So long as dispersion is negligible, the electromagnetic wave generated by this source is described by a vector potential
 \beq \bA_s(t,\br) = \int d^3\br' \fr{1}{4\pi |\br-\br'|} \bj_s^\text{T}(t-|\br-\br'|,\br'), \label{jtoA}\eeq
where $\bj_s^\text{T}$ is the transverse (divergenceless) part of the current $\bj_s$.
%It is important to note however that in the quantum description each such pulse \emph{does not} correspond to one photon.  Some sources may remain coherent for longer and emit several photons coherently.  Very transitory ones might emit on average less than one photon (of course, that means sometimes one, other times none).
The question to be resolved in due course is the correspondence between pulses and photons, as far as the distant observer is concerned.
The precise form that these functions $\bj_s$ take is of no importance.  In estimating the results, we have to average over the time around which it is centered, over the phase of the emitted waves and over the various parameters that describe the shape of $\bj_s$.

But before doing that, we can simplify the picture.  We are interested only in the value of the wave in the vicinity of the detector, placed at say $\br=(0,0,z)$ where always $z$ is much larger than the values of the coordinates of $\br'$ in (\ref{jtoA}).  So it is possible to replace $|\br-\br'|$ by $(z-z')$ in the time argument of $\bj_s$, and by $z$ is the much more slowly varying denominator.  This also means that transverse components of $\br$ are irrelevant, so the value of $\bA$ is effectively constant across the detector.  In other words, we are really only talking about plane wave propagation along the $z$ axis, described by the two functions $A_x(t,z)$ and $A_y(t,z)$.  To simplify even further, we can suppose that a polarizer is introduced in front of the detector to eliminate one of the two plane polarizations, leaving only one amplitude function, say $\cA=A_x$.  Consequently, (\ref{jtoA}) reduces to an expression of the form
 \beq \cA_s(t,z) = \cA^{(+)}_s (t,z) = \fr{1}{4\pi z} \int d^3\br'\, j_{s,x}^\text{T}(t-z+z',\br') = \fr{F_s(t-z-t_s)}{4\pi z} e^{-i[\om_0 (t-t_s) + \vp_s]},
 \label{source}\eeq
where again we assumed a dominant frequency $\om_0$.

We could now treat $z$ as a constant for the source (or observer) position, and reasonably choose to model $F_s$ by a function of generic form,  namely \beq \fr{F_s (t-z-t_s)}{4\pi z} = f (t-t_s). \label{gfs} \eeq Here we adopt the commonly used model of a Gaussian pulse \beq f(t) = a e^{-t^2/(2\si^2)}. \label{fs} \eeq  Hence, with \beq \cA^{(+)} (t,z) = \sum_s \cA^{(+)}_s (t,z), \label{A} \eeq we have the ensemble average \beq \<\cA^{(-)} (t) \cA^{(+)} (t')\> = \sqrt{\pi} a^2 \lambda\si e^{-(t-t')^2/(4\si^2)} e^{i\om_0 (t-t')}, \label{AA} \eeq  where $\lambda$ is the number of pulses per unit time; (\ref{AA}) is a consequence of the random uncorrelated phases $\vp_s$ among the sources, which means only terms with $s=s'$ in the double summation leading to (\ref{AA}) survive the averaging.  The summation over $s$ in (\ref{A}) is also the reason why $t_s$ no longer appears in $t_s$.

%To calculate other first order quantities, we start with $\<\hat M (t)\>$.

\subsection{The detector}

Consider a simple photoelectric detector, of area $S$, which records the number of photons arriving during any particular time interval.  For simplicity assume an efficiency factor $\et=1$.
%( smaller value of $\et$ can be accommodated by introducing a beam splitter in front of the perfectly efficient detector.)
The detector is preceded by a frequency filter that transmits a narrow band of frequencies of width $\de\om$ centered on $\om_0$, a polarizer, and a telescope that transmits only light from a very small solid angle.  The field at the detector is thus described by a vector potential $\cA(t,z)$ which is a sum of all the terms of the form (\ref{source}), if there is no dispersion of the light by the intervening medium.

Now in a semiclassical treatment, the number of photons detected within the time interval $\De t$ is
 \beq N(t_0,\De t) = \int_{t_0}^{t_0+\De t} dt\,M(t), \label{Nt0} \eeq
where $M$ is given by the semiclassical equivalent of (\ref{M}), namely
 \beq M(t) = \al^*(t)\al(t), \qquad \text{with} \qquad \al(t)=\int\fr{d\om}{\sqrt{2\pi}}\al(\om) e^{-i\om t}, \label{Mal} \eeq and $\al^* (t)$ given by the conjugate of this equation.
We are interested in the statistical properties of this rate.  Of course, talking about `photons' in a classical context is a little odd.  This quantity $N$ here is not in any way restricted to being an integer.  Clearly this description cannot be entirely correct; $N$ can at best be the \emph{expectation value} of the number of photons detected.  We return below to a proper quantum treatment.  From (\ref{MAA}) and (\ref{Mal}), however, we obtain the semi-classical result, \beq M(t) = \al^* (t) \al (t) = 2S\om_0 \cA^{(-)} (t) \cA^{(+)} (t). \label{Mt} \eeq The ensemble average is, from (\ref{AA}), \beq \< M(t)\> = \<\al^* (t) \al (t)\> = 2\sqrt{\pi} S\om_0 a^2 \lambda\si, \label{Mavg} \eeq
where the last equality is valid only under the scenario of a distant source emitting many identical Gaussian pulses with random (carrier wave) phases and at random times, at the mean rate of $\lambda$.

We also wish to compute the averages $\<\al^* (t) \al (t')\>$ and $\<\al^* (\om)\al (\om')\>$. From (\ref{AA}) and (\ref{Mal}b), one finds \beq G(t-t') = \<\al^* (t) \al (t')\> = 2\sqrt{\pi} S\om_0 a^2 \lam\si e^{-(t-t')^2/(4\si^2)} e^{i\om_0 (t-t')}. \label{Gt} \eeq  Thus $G(t)$ has a width $\ap\si$.   More generally,
the function $G(t)$ is given by \beq G(t) = 2S\om_0 \lam\int dt' g (t')g^* (t'-t),~{\rm where}~g(t) = f(t) e^{-i\om_0 t} \label{gGt} \eeq and $f(t)$ as defined in (\ref{gfs}) does not have to involve a Gaussian; rather, any real function $f(t)$ for the envelope would suffice, and $G(t)$ will have the same width as this envelope.  In particular, however, \beq \<M(t)\> = G(0).  \label{MG0} \eeq If we further define $\tilde G(\om)$ as  \beq \tilde G(\om) = \int dt~G(t) e^{i\om t}, \label{Gw} \eeq and enlist (\ref{Mal}b) again, we obtain \beq \<\al^* (\om)\al (\om')\> = \de (\om-\om') \tilde G(\om). \label{al*walw'} \eeq  In general, \beq \tilde G(\om) = 2S\om_0 \lam |\tilde g(\om)|^2 \label{gGw} \eeq where $\tilde g(\om)$ is the Fourier transform of $g(t)$.
%Both (\ref{Gt}) and (\ref{al*walw'})
All these features
are consequences of our model of the distant source as a statistically stationary emitter, namely one with no preferred zero of time, or (equivalently) the constituent Fourier components of $\cA(t)$ are uncorrelated.  In the case of $t=t'$ the first equation of (\ref{Gt}) gives, with the help of (\ref{Mavg}), \beq \fr{1}{2\pi} \int\tilde G(\om) d\om = G(0) = \< M(t)\> = 2\sqrt{\pi} S \om_0 a^2\lam\si, \label{intspec} \eeq a result that not only enforces the Parseval theorem, but also indicates that $\tilde G(\om)$ is the spectrum of the beam.  The width $\de\om$ of the spectrum and the width $\si$ of $G(t)$ obeys the relation \beq \de\om \ap \fr{1}{\si} \label{Hei} \eeq by virtue of the fact that $G(t)$ and $\tilde G(\om)$ are Fourier pairs.

Note that while $\tilde G(\om)$ is real, this is not true of $G(t)$, which by (\ref{Gw}) satisfies the relation
 \beq G^*(t) = G(-t). \label{G*G} \eeq
From (\ref{intspec}), the \emph{average} counting rate $G(0)$ must as a result of (\ref{G*G}) be real.  Hence, for an exposure time of $T$ to the beam,
 \beq \<N(t_0,T)\> = G(0) T. \label{Nav}\eeq
We shall see that dispersion only changes the phase of each Fourier component, but not the value of $G(t)$; in particular, this means the average counting rate $G(0)$ is unaffected by the intervening medium.

\subsection{Semiclassical treatment of brightness fluctuations}

To evaluate the variance of the photon count number, we need to examine the second-order correlation function
\beq \<M(t)M(t')\> =  \<\al^*(t)\al(t)\al^*(t')\al(t')\>. \label{corr2} \eeq  From (\ref{Mt}) and (\ref{A}), one sees that as a result of the random phases $\vp_s$ there are two types of surviving terms: those from pairwise correlations in which there are only two distinct pulses $s,s'~(s\neq s')$ among the four participating ones, and those involving $s=s'$ (\ie~all four pulses are the same).  In general these two terms are written down in the same order as follows, \beq \<M(t)M(t')\> - \<M\>^2 = |G(t-t')|^2 + 4S^2 \om_0^2 \lam \int dt'' |g(t+t'')|^2 |g(t'+t'')|^2 \label{gshoot} \eeq where $g(t)$ was defined in (\ref{gGt}).

For Gaussian pulses in particular, the $|G(t-t')|^2$ term leads to a $\lambda^2\si^2$ contribution to the two-point function, while the latter $\lam\si$ which is much smaller in the $\lambda\si \gg 1$ limit of many overlapping pulses.  Thus, \beq \<M(t)M(t')\> - \<M(t)\>^2 =  \left[S^2\om_0^2 a^4 \lam\si e^{-(t-t')^2/(2\si^2)}\right] (4\pi\lam\si + 2\sqrt{2\pi}). \label{MtMt'} \eeq  For large $\lam\si$ one discards the last term, and \beq \<M(t)M(t')\> - \<M(t)\>^2 \ap |G(t-t')|^2,~\lam\si \gg 1. \label{MMapprox} \eeq  which yields the radiometer equation of radio astronomy as a corollary, as we demonstrate immediately below.

%Now each of the factors here is a sum over a very large number of terms, the number of distinct sources that can contribute to the field at the detector.  This is effectively the number of distinct sources that are active at a given time.  We could estimate the number from a knowledge of the luminosity of the source, but this is hardly worth doing.  It is not important to know precisely how large the number is, only that it is clearly \emph{very much} larger than unity.
%Because the phases of the different sources are random and uncorrelated, when we substitute the sums over $s$ into (\ref{corr2}) we need only keep terms in which the phases cancel out.  The only significant contributions are those where each $\al$ is paired with an $\al^*$ from the same source.  There is in principle a contribution from terms in which all four factors come from the same source, but this contribution is clearly tiny, and we get simply
% \beq \<M(t)M(t')\> = G(0)^2+|G(t-t')|^2. \label{MMav} \eeq
%The important thing about this expression is that it is unaffected by dispersion, since $G(t)$ is.  Correspondingly, for

The variance of photon counts over some exposure time $T$ may now be derived from the formula
\beq (\de N)^2 = 2\int_0^T dt\,(T-t) [M(t)M(0)-\<M\>^2] \label{DeltaN2}\eeq which follows from squaring (\ref{Nt0}) and taking the ensemble average.  Under the $\lam\si \gg 1$ scenario, one may examine the behavior of the variance as $T$ becomes very large or very small.  In the former case, the integral is dominated by the region where $t\ll T$, \ie~
 \beq (\de N)^2 \ap 2T \int_0^\infty dt\, |G(t)|^2. \eeq
We may then write
 \beq (\de N)^2 \ap \fr{\de t}{T}\<N\>^2\qquad (\de\om \De t \gg 1), \label{dN} \eeq
where $\de t$ is given by
 \beq \de t = \fr{1}{G(0)^2} \int_{-\infty}^{\infty} |G(t)|^2 dt, \eeq
and by (\ref{Gt}) is of order $\si \ap 1/\de\om$ with $\de\om$ being the width of $\tilde G(\om)$, which is essentially the width of the spectrum of the beam.  Thus, an alternative form of (\ref{dN}) is \beq \left(\fr{\de N}{\bar N}\right)^2 = \fr{1}{T\de\om},~{\rm for}~T\de\om\gg 1, \label{radiometer} \eeq which is the radiometer equation.

In the opposite limit of $T\de\om \ll 1$, another approximation scheme should be applied to (\ref{DeltaN2}), namely $G(t)$ is nearly constant within the range of integration, equal to $G(0)$.  We find
\beq (\de N)^2 \ap \<N\>^2\qquad ,~{\rm for}~T\de\om\ll 1. \eeq  Together with (\ref{radiometer}), the behavior of $(\de N)^2$ indicates that the maximum relative variance is reached at $T\de\om \ap 1$.  When $T$ becomes even smaller, the variance no longer increases like (\ref{radiometer}) because the fluctuations, sometimes known as photon bunching noise, are correlated on timescales beneath the coherence length of $\si\ap 1/\de\om$.

A major inadequacy of the present treatment of fluctuations is the omission of Poisson noise, also referred to as the shot noise, which has the characteristic of $(\de N)^2 = N$ for {\it all} exposure times $T$ however large or small.  This indicates shot noise is a point process that originates from the corpuscular (particle) nature of light, \ie~it is a consequence of the quantum nature of radiation, and cannot emerge from the semiclassical formalism here.

\subsection{Quantum treatment of brightness fluctuations}

The first question here is how to represent the beam emanating from the source, but there is a very simple and well-established answer to that.  The effect of classical current sources on the quantum vacuum state is to create coherent states.  So it is natural to assume that the wave emanating from the source in our problem is represented by this coherent state $|\cA\>$.  Then all the calculations above go through more or less as before.

The only change that must be made is that when we write down the expression for the photon counting rate, we have to subtract the formally infinite vacuum value.  In other words, we have to write the counting rate operator in normally ordered form, (\ref{M}).  Explicitly, since
 \beq \<\cA|:\hat M(t):|\cA\> = \<\cA|\hat a^*(t)\hat a(t)|\cA\> = \al^*(t)\al(t), \label{norder} \eeq
this means that the value of average photon number count $\<N(t_0,T)\>$, given by (\ref{Nav}), is finite.

However, the situation is quite different for the second-order correlation function.  From (\ref{atcomm}) and (\ref{M}), we find
 \beq :\hat M(t) \hat M(t'):~= \hat a^*(t)\hat a^*(t')\hat a(t)\hat a(t') = \hat M(t) \hat M(t') - \hat a^*(t)\hat a(t')\de(t-t') = \hat M(t) \hat M(t') - :\hat M(t):\de(t-t'). \label{MMcorr} \eeq
%\hat M(t) \hat M(t') = \hat a^*(t)[\hat a^*(t')\hat a(t)+\de(t-t')]\hat a(t')
%=\ :\hat M(t) \hat M(t'): +\ \de(t-t')\hat M(t). \label{MMcorr} \eeq
Now the ensemble average of the expectation value of the normally-ordered product has precisely the same value (\ref{MMapprox}) as the semiclassical treatment in the limit of infinite $\lam\si$, namely (after applying (\ref{MG0}) with $M(t)$ replaced by $:M(t):$),
 \beq \<:M(t) M(t'):\> = [G (0)]^2+|G(t-t')|^2. \eeq
We thus obtain,
 \beq \<M(t) M(t')\> - \<:M(t):\>^2
 = |G(t-t')|^2 + G(0)\de(t-t'). \label{MM} \eeq
If we then apply (\ref{DeltaN2}) and
integrate over $t$, we find for the variance of the photon number count
 \beq (\de N)^2 = 2\int_0^T dt\,(T-t) |G(t)|^2 + N,~{\rm for~{\it any}}~T, \label{shot} \eeq
where the last term equals $G(0)T$.
This is identical to the semiclassical result (\ref{DeltaN2}) except for the addition of the final term, which is now the Poisson shot noise.  Note that this extra term is simply $\<N\>$, as is to be expected of shot noise.

The picture is now very similar to the way in which \cite{hir14} derived the shot noise term, and highlights the origin of this term as the particle nature of radiation, namely photons.  In this approach, the pulse arrival rate $\lam$ is so enormous that the last term of (\ref{gshoot}) is negligible, it certainly has nothing to do with photon shot noise.  Nevertheless, the reason why we said before (\ref{MMcorr}) that the derivation of the first and second order quantities differ has to do with normal ordering.  On one hand, when evaluating the mean count rate $\<M(t)\>$, one equated it to the ensemble average of $\<\cA|:\hat M(t):|\cA\>$ to remove an infinity, see (\ref{norder}).  On the other hand, for $\<M(t) M(t')\>$ one equated it to the ensemble average of $\<\cA|\hat M(t) \hat M(t')|\cA\>$, \ie~{\it without} normal ordering.  Thus there is a basic inconsistency in this formalism for computing measurable quantities, even if it leads to agreement with observations.

The alternative approach to the shot noise problem, provided in \cite{lieu13}, does not suffer from the same difficulty.  Here we wish to revisit it to highlight the point, and also to show that \cite{lieu13} does nevertheless have its own unique problems in explaining certain other aspects of the behavior of radiation -- problems not shared by the aforementioned approach (which is also the generally accepted one).  To be precise, we are not referring to the difference between \cite{hir14} and \cite{lieu13} in their prediction of line-of-sight dispersion effects (since we have not addressed the phenomenon of dispersion yet); rather, there are other effects that can be already used to experimentally compare and contrast these two models.

In \cite{lieu13}, the shot noise component is attributed to the $\lam\si$ term of (\ref{MtMt'}), in the sense that $\lam$ is treated as the photon arrival rate at the observer, \ie~each photon is a pulse.  This assumption greatly limits $\lam$ to much more modest values such that the last term of (\ref{MtMt'}) cannot always be ignored\footnote{This is quite unlike \cite{hir14}, which considers each spherical wave packet emitted by an atom or ion to be a pulse the observer has finite probability of detecting.  Since there are many emitting atoms or ions in the quasar, each pulse has only a tiny fraction of a photon's energy, \ie~the modulus square of the sum of the many overlapping (interfering) pulse amplitudes gives the probability of an arriving photon.  Thus $\lam$ for the spherical atomic pulses is far larger than the photon pulses, but experiments clearly preferred the former as the correct model.}  To elaborate, (\ref{Mavg}) and (\ref{MtMt'}) together imply \beq \<M(t)M(0)\> - \<M\>^2 = |G(t)|^2 + 2\sqrt{2\pi}S^2\om_0^2 A^4 \lam\si e^{-t^2/(2\si^2)}.  \label{shoot} \eeq    Substituting (\ref{shoot}) into (\ref{DeltaN2}), one again obtains (\ref{shot}), but {\it only} in the limit of $T \gg \si$, or equivalently $T\de\om \gg 1$.  The reader can verify that unlike (\ref{shot}), which holds for all $T$, the expression for $(\de N)^2$ here no longer has $N$ as its last term, rather some other quantity less than $N$, when  $T\de\om < 1$.  To understand the discrepancy, one must compare the last terms of (\ref{MM}) and (\ref{shoot}).  In the former, the term is a Dirac delta function that portrays the impossibility of `resolving a photon', \ie~shot noise obeys Poisson statistics from indefinitely large to infinitesimally small scales.  In the latter, the point process of shot noise is manifested only on long timescales, but not on scales small enough to probe the exponential structure of the last term of (\ref{shoot}).

Turning to experimental evidence, the cross correlation at zero lag between the intensity time series of two beams that emerge from a 50:50 beam splitter when the incident beam consists of incoherent radiation, \cite{han57}, revealed only the first term\footnote{One should not be surprised at all that the first term was detected in the cross correlation, because its origin is the genuine wave nature of light.  Thus intensity fluctuations caused by this term are found having identical patterns in the two beams that emerged from the beam splitter from the same primary (incident) beam.}  of (\ref{MM}) but not the last term, see \eg~the discussion on the $g^{(2)}$ correlator in \cite{lou00}.  This clearly demonstrates that photons are fundamentally structureless and indivisible at the beam splitter, \ie~a photon can randomly be assigned to one of the two emerging beams, but cannot be in both beams.  Thus the evidence is consistent with (\ref{MM}).  In (\ref{shoot}), however, the last term also owes its origin to the wave nature of light in the sense that pulses (or wave packets) can readily be divided by the beam splitter.  This means, according to (\ref{shoot}), the cross correlation should contain a contribution from the last term, and both terms should exhibit correlation power at finite lag, provide the lag is not $\gg \si$.  In reality, one only sees the first term at zero and small lags.  Moreover, the fact that even the cross correlation at zero lag exhibited no sign of the shot noise contribution irrespective of the timing accuracy of one's apparatus lends further support to the absence of a scale that marks the departure of photon shot noise from genuine Poisson statistics.

Thus the weight of observational evidence appears to strongly favor (\ref{shot}), namely the view that the distant light source consists of an almost infinite number of atoms and electrons emitting spherical pulses of radiation all distant observers can see (hence the pulse rate $\lam \to\infty$, reducing the relative importance of the last term of (\ref{gshoot}) to zero), so photon shot noise cannot be due to pulse fluctuations.  Rather, it is a consequence of the normal ordering recipe of quantum operators, even if this recipe is not self-consistent in the sense that one calculates the mean photon rate and its variance differently.  Regrettably there are no further options: one has to settle the matter somewhat unsatisfactorily in this fashion.  Only when we do that, will it be possible to understand how \cite{lieu13} wrongly predicted the effect of dispersion on incoherent light.

\subsection{Dispersion}

The effect of dispersion is already discussed in detail in \cite{lieu13}, but see also \cite{boh51}, and \cite{bor70}.  In essence, if a beam with characteristics of (\ref{source}) and (\ref{fs}) propagated through a uniform dispersive medium, it will arrive with the source function modified to \beq \cA_s(t,z) = \cA^{(+)}_s (t,z) = e^{-i\vp_s} \int \fr{d\om}{2\pi} \tilde g(\om) e^{i\om t_s}
e^{-i\om t + i\om_0 z/v_p + i(\om-\om_0)z/v_g + i k_0^{''} (\om-\om_0)^2 z/2 + \cdots}, \label{dispersed}\eeq
where $\tilde g(\om)$ is the Fourier transform of $g(t)$ with $g(t)$ as defined in
(\ref{gGt}),
the phase and group velocities are \beq v_p = \fr{\om_0}{k_0} = \left(1-\fr{\om_p^2}{\om_0^2}\right)^{-1/2} \ap 1+\fr{\om_p^2}{2\om_0^2}, \label{vp} \eeq
and \beq v_g = \left(\fr{d\om}{dk}\right)_{k=k_0} = \left(1-\fr{\om_p^2}{\om_0^2}\right)^{1/2} \ap 1-\fr{\om_p^2}{2\om_0^2}; \label{vg} \eeq respectively, and $k^{''}_0 = (d^2 k/d\om^2)_{\om_0}$.

The crucial point to emphasize is that dispersion does not change the functions $\tilde G(\om)$ and $G(t)$ as defined in (\ref{gGw}) and (\ref{gGt}) respectively.  To see this, we first
observe that the integrand of (\ref{dispersed}) less the $e^{-i\om t}$ factor is in fact the Fourier amplitude of the original pulse $f(t-t_s) e^{-i\om_0 (t-t_s)}$, namely $\tilde g (\om) e^{i\om t_s}$, {\it after} this amplitude is modified by the passage of the beam through the dispersive plasma column.  Thus the effect of dispersion is to replace $\tilde g(\om)$ by
$\tilde g' (\om)$, \ie~ \beq  \tilde g(\om) \to \tilde g' (\om) =  \tilde g(\om) e^{i\om_0 z/v_p + i(\om-\om_0)z/v_g + i k_0^{''} (\om-\om_0)^2 z/2 + \cdots}. \label{tgprime} \eeq  Yet, from (\ref{gGw}), the beam spectrum $\tilde G (\om)$ does {\it not} depend on the phase of $g$ (now $g'$).  This means, $G(t)$ is unchanged by dispersion.  It then follows from (\ref{MG0}) and (\ref{MM}) that the mean photon rate and its autocorrelation function, \ie~both first and second order quantities, are also unaffected.  More generally, \cite{wan89} proved that, if shot noise is ignored, all $n$-point functions of the photon rate (or the intensity) are unaffected by dispersion.  Fundamentally, the reason has to do with the random uncorrelated phase $\vp_s$ of the pulses.  While dispersion adds a systematic phase $\xi_s$ to $\vp_s$, there is no net effect because it does not alter the fact that once $\<e^{i\vp_s} e^{-i\vp_{s'}}\>$ vanishes,  $\<e^{i(\xi_s +\vp_s)}e^{-i(\xi_{s'} + \vp_{s'})}\>$ must also vanish (assuming $s \neq s'$ in both cases).

The conclusion is that \cite{hir14} were right in asserting the absence of any measurable imprints, of the sort described in \cite{lieu13}, due to the passage of incoherent light through a dispersive medium.  It should also be pointed out that if the shot noise model were that of \cite{lieu13}, then the last term of (\ref{shoot}) or (\ref{gshoot}) would indeed have represented observable evidence of dispersion because of pulse broadening.  But, as already explained in the last section, even `non-dispersive' experiments on incoherent light do not support the \cite{lieu13} model of photon shot noise.  It should therefore come as no surprise that the analysis of radio quasar data by \cite{hal16} failed to reveal the dispersive effects predicted by the two 2013 papers of Lieu.

%obtained by the standard procedure.  We Fourier transform $f_s$ to get
% \beq \tilde f_s(\om) = \int dt\,f_s(t)e^{i\om t}. \eeq
%We are going to use a detector with a narrow-band filter, centred on a frequency $\om_0$, so we are really only interested in the dispersion relation in that vicinity.  Let us assume that near $\om_0$ the dispersion is well represented by the quadratic formula
% \beq k_\om = \om_0/c + \xi(\om-\om_0) + \beta(\om-\om_0)^2, \eeq
%where $c$ is the speed of a monochromatic wave of frequency $\om_0$, and $\xi$ and $\beta$ are %constants.  Then in place of (\ref{source}) we have
% \beq f_s(t,z) = \int \fr{d\om}{2\pi} \tilde f_s(\om)
% e^{-i\om t + i\om_0 z/c + i\xi(\om-\om_0)z+i\beta(\om-\om_0)^2 z}. \label{dispersed}\eeq

%\beq \left(\fr{\de N}{\bar N}\right)^2 = \fr{\bar n^2 + \bar n}{\sqrt{2\pi} \lam^2 \si T}. \eeq

%This is still expressed in terms of $G(t)$ and so is unaffected by dispersion.

%The overall conclusion is that there will be no observable effect of dispersion.

\section{Is there really no imprint of dispersion on stationary light?}

In the foregoing sections we extended the formal treatment of \cite{wan89} to include the particle (photon) nature of light, thereby confirming the assessment of \cite{hir14} that the autocorrelation techniques proposed by \cite{lieu13} cannot be used to infer the dispersion of light from relatively steady incoherent sources.  Now we turn to the question whether this means dispersion of stationary light leaves behind any tangible and measurable imprint at all.

Given the development of the previous sections, there is actually a much simpler way of looking at the whole problem.  Provided that the pulse and photon arrival rates satisfy $\lam\si \ggg 1$ and $\<M\> \si \gg 1$ (the latter means the photon occupation number of the beam is $\gg 1$), conditions that are easily fulfilled in radio astronomy, one can ignore any pulses and photons in the beam, by modeling it as genuinely stationary light comprising many modes with random phases.  Within some narrow band centered at frequency $\om_0$ and of bandwidth $\de\om \ll \om_0$, therefore, one could depict a beam propagating through vacuum as \beq \cA_s(t,z) = \cA^{(+)}_s (t,z) =  \fr{1}{T} \sum_s \tilde h(\om_s)
e^{-i\om (t-z)+i\vp_s}  \label{A+s} \eeq where we assume \beq h(\om_s) = h_0 e^{-(\om_s-\om_0)^2/2\si^2} \label{h} \eeq is the spectral amplitude at $\om_s$ as enforced by a passband filter or spectral line of the source, and an integral over frequencies of the form (\ref{dispersed}) is replaced here by a sum over modes, noting that the bandwidth $\de\om$ obeys (\ref{Hei}) and (from the periodic boundary condition as applied over the time span $T$) the number of modes is $\ap T/\si$ where $T$ is the total exposure time.  In this way the mean photon arrival rate at a fixed position $z$ of the observer and for various $t$ is again given by (\ref{Mt}) and (\ref{MG0}), and the two-point function by (\ref{MMapprox}), with the quantities $\cA^{(+)} (t)$ and $\<\cA^{(-)} (t) \cA^{(+)} (t')\>$ given by (\ref{A}) and (\ref{AA}) respectively.

If the beam has passed through a dispersive medium, however, (\ref{A+s}) will be modified to become \beq \cA_s(t,z) = \cA^{(+)}_s (t,z) =  \fr{1}{T} \sum_s \tilde h(\om_s) e^{-i\om t + i\om_0 z/v_p + i(\om_s-\om_0)z/v_g + i k_0^{''} (\om_s-\om_0)^2 z/2 + \cdots}, \label{A+sd} \eeq
%at the observer's position $z$ and for various $t$,
with $v_p$, $v_g$, and $k_0^{''}$ as defined in and around (\ref{vp}) and (\ref{vg}).  The rest of the situation as already described in the previous subsection, namely $G(t)$ and $\tilde G(\om)$ are unchanged by dispersion and there can be no imprint in such a context.  Physically, this means the modes of chaotic (\eg~thermal) light are phase uncorrelated, \ie~the total brightness is proportional to the sum of the occupation number of each mode, or equivalently the sum of the modulus squares of the mode amplitudes.  A `cross term', or product of two different mode amplitudes, does not contribute because $\<\exp i(\phi_s - \phi_{s'})\> = 0$ when the two mode phases are random and uncorrelated, and dispersion simply adds a systematic phase to $\phi_s$ which does not alter the above picture of mode independence.  Yet this does not mean, if each mode phase $\phi_s$ comprises a random component $\vp_s$ and a systematic quadratic phase $k_0^{''} (\om_s-\om_0)^2 z/2$ that is imprinted by the dispersive medium, the latter phase is necessarily masked by the former from detection.  In fact, we shall show how an analysis of phase information from an ensemble of individual modes that is more sophisticated than simple amplitude cross products can reveal the quadratic phase.  This is a reasonable conclusion, since phase and occupation number are separate observables after all (quantum mechanically they are non-commuting dynamic variables).

As a simple illustration, lets us remove all inessentials by writing the phase of one mode of chaotic light with and without passage through a dispersive medium, as \beq \phi_s = \vp_s + \tfrac{1}{2} k_0^{''} (\om_s-\om_0)^2 z, \label{1mode} \eeq and $\phi'_s = \vp'_s$ respectively, with $\vp_s$ and $\vp'_s$ being random uncorrelated phases.  If the former comes from a test beam and the latter some reference beam in the laboratory, any interference effect between the two beams will involve a term $\propto \<{\rm Re}~[\exp i(\phi_s - \phi_{s'})]\> = \<\cos [\vp_s - \vp'_s + k_0^{''} (\om_s-\om_0)^2 z/2]\> =0$, and the same conclusion would apply even in the absence of the quadratic phase; in this respect the test beam responds in the same way whether there was dispersive propagation or not.  But, on the other hand, if one {\it directly} measures the electric oscillation of many individual modes  to determine their phases $\{\phi_s\}$ and $\{\phi_s'\}$ at some fixed time epoch, as can readily be done with available technology (see below), then the ensemble average of these phases {\it themselves} will differ: $\<\phi'_s\> = \<\vp'_s\> = 0$, whereas $\<\phi_s\> = \<\vp_s\> + \<k_0^{''} (\om_s-\om_0)^2 z/2\> = k_0^{''} (\om_s-\om_0)^2 z/2$ does not vanish.  In fact, $\<\phi_s\>$ is frequency dependent, and from this dependence as ascertained by examining many modes one can in principle deduce the value of $k_0^{''}$, hence the column density of dispersion, as we shall argue in detail.

Let us therefore write the Fourier amplitude of a model in full {\it viz.}~\beq \tilde\cA^{(+)}_s (\om) = \fr{1}{T} \tilde h(\om_s) e^{i\vp_s+i\om_0 z/v_p + i(\om_s-\om_0)z/v_g + i k_0^{''} (\om_s-\om_0)^2 z/2 + \cdots} \label{A+sw} \eeq where the
%The validity criterion of (\ref{A+sw})
%are the absence of frequency redshift, and
%is the negligibility of higher order terms than $k_0^{''}$.
%The first means we must confine our discussion to Galactic sources and the interstellar medium, and treat the problem of quasars and the intergalactic medium in a future work.  The second usually holds well for both inside the Milky Way and beyond.  Thus, for the Galactic interstellar medium with free electron density $n_e \ap 10^{-3}$~cm$^{-3}$, \cite{wel09}, the
$k_0^{''}$ term of the exponent is, in terms of the typical density of the intergalactic medium and distance to quasars, \beq \tfrac{1}{2} k_0^{''} (\om-\om_0)^2 z = 26.5 \left(\frac{n_e}{10^{-7}~{\rm cm}^{-3}}\right) \left(\fr{\om_0}{6~{\rm GHz}}\right)^{-1} \left(\fr{\de\om/\om_0}{10^{-4}}\right)^2 \left(\fr{z}{1~{\rm Gpc}}\right), \label{kpp} \eeq and with the chosen bandwidth $\de\om = 0.6$~MHz the next order term is $\ap\de\om/\om_0 = 10^{-4}$ times the $k_0^{''}$ term, which puts it at $\ll 1$ values.

%\beq \tfrac{1}{2} k_0^{''} (\om-\om_0)^2 z = 26.5 \left(\frac{n_e}{10^{-3}~{\rm cm}^{-3}}\right) \left(\fr{\om_0}{6~{\rm GHz}}\right)^{-1} \left(\fr{\de\om/\om_0}{10^{-3}}\right)^2 \left(\fr{z}{1~{\rm kpc}}\right), \label{kpp} \eeq and the next order term is $\ap\de\om/\om_0 = 10^{-3}$ times the $k_0^{''}$ term, which puts it at $\ll 1$ values.

Now the question naturally arises: can the individual modes be measured to probe the quadratic spectral phase of dispersion, namely the $k_0^{''}$ term?  SETI observers have demonstrated $1$~Hz spectral resolution at GHz frequencies, provided the total exposure $T$ is long enough to reach the resolution whilst ensuring the detection of at least one mode, \cite{har16}.  For an exposure time of merely $T\ap 6$~s,
%and assuming $\de\om = 0.6$~MHz ($\de\nu\ap~1$~MHz) at $\om_0 = 6$~GHz ($\nu_0\ap 1$~GHz),
the number of modes is \beq N_{\rm mode} \ap T/\si = \fr{T\de\om}{(2\pi)},
\ap 6 \times 10^6,
\label{Nmode} \eeq \ie~the spacing between modes is indeed $d\om = 2\pi/T \ap 1$~Hz. Thus, longer exposure means even better spectral resolution.  Moreover, if one digitizes an incoherent detector time series (waveform) spanning the total duration $T$ at a factor $> 10^3$ times beneath the coherence time $\si\ap 1/\de\om$, one would directly sample the electric field with the help of an amplifier to measure the phase of the mode (relative to the some reference which is the same for all modes), as is done in radio astronomy at low frequencies anyway.  Since the assumption is that the source emission has high occupation number, this would not add noise unless the amplifier noise temperature were higher than the source temperature.

If such phase measurements are simultaneously made in a single exposure of $T$ at various frequencies located {\it symmetrically} on either side of $\om_0$ across the full Gaussian bandwidth of (\ref{h}), the data would consist of a set of phases $\{\phi_s,~s=1,2,\cdots, n\}$ where $n\leq N_{\rm mode}$ can be as large as $6\times 10^5$ in just 6~s from the previous paragraph, and  \beq \phi_s =  \vp_s+\fr{\om_0 z}{v_p} + \fr{(\om_s-\om_0)z}{v_g} + \tfrac{1}{2} k_0^{''} (\om_s-\om_0)^2 z + \cdots. \label{phis} \eeq  Computing the sample mean of the phases, the first and third terms vanish (the former due to mode phase decorrelation, the latter by the symmetric distribution of $\om_s$ about $\om_0$), and the second term is a constant equal to $\om_0 z$ if we assume (without introducing any appreciable error) that the phase velocity is $v_p \ap c=1$.  Thus \beq \bar{\phi}_s = \tfrac{1}{2} k_0^{''} z \overline{(\om_s-\om_0)^2}+~{\rm const.} = \tfrac{1}{2} k_0^{''} z\si^2+~{\rm const.} \label{barphi} \eeq  Although the last term is a very big constant, it should not act as a deterrent because there is always a reference phase value to be subtracted, and in any case the convention of phase measurements periodically wraps $\phi_s$ around the range $-\pi \leq \phi_s \leq \pi$.

Thus the strategy here is to fine tune the addition or subtraction of a large constant to the measured $\bar\phi_s$ such that the resulting value becomes proportional to $\si^2$, the constant of proportionality then yields the sought after information on $k_0^{''}$, hence the dispersive column density $n_e x$ of the intergalactic medium via (\ref{kpp}).  This means one has to repeat the measurement with various spectral filtering, namely different $\si \ap \de\om$ but the same central frequency $\om_0$, until one successfully detects the $k_0{''}$ term.  The difficulty of following through this proposed idea lies in the stability of the frequencies at which $\phi_s$ is sampled, especially the central frequency $\om_0$ because even the slightest fluctuation in $\om_0$ could trigger a very large phase error in the constant, \ie~$\om_0 z$, term of (\ref{barphi}).  This is a challenge on the absolute stability of one's clock which, as we shall see below, can only marginally be met with current technology.

The problem is completely mitigated here, however, by the {\it simultaneous} sampling within the same exposure time $T$ of the phase $\phi_s$ of as many of the $N_{\rm mode}$ available modes as possible, where $N_{\rm mode}$ is given by (\ref{Nmode}).   This can be achieved by sampling the received waveform sufficiently fast for sufficiently long time $T$ and then performing fast Fourier transform (FFT) to the sampled data.   As already noted immediately after (\ref{Nmode}), the total sampling duration determines the frequency spacing of these modes.  The FFT produces not only the amplitude of each mode but also their relative phases referenced to the carrier wave $\om=\om_0$ at some time epoch. The latter is exactly what we need here.

Provided the modes are distributed symmetrically across the spectral filter in frequency space, one can take the phase average of a smaller subset of the full sample, starting with the core of the filter around $\om_0$, then moving outward to cover pairs of equally-sized frequency interval placed symmetrically away (\ie~equidistant) from the core region of the Gaussian.  In this way the averages $\bar\phi_s$ will all be evaluated using exactly the same $\om_0$ because of the simultaneity of the individual measurements of $\phi_s$, but increasing $\si$ as one subset moves further away from $\om=\om_0$ than the previous.  Since, as explained after (\ref{Nmode}), one has up to $6 \times 10^5$ data points for $\phi_s$ over a mere 6~s interval, and a more realistic exposure time for radio telescopes is much longer than 6~s, there should be no lack of samples for the purpose of averaging over each subset.  Beware however that the value of $\si$ for each subset is to be calculated by assigning equal weights to the relevant frequencies and not in accordance with $\tilde h(\om_s)$, because within the region of appreciable $\tilde h (\om_s)$ where phase measurements can be made, the phase of a mode is independent of its amplitude, and participates in (\ref{barphi}) equally as any other in the same subset.

Another possible snag in this approach pertains to its requirement of symmetrical distribution of the modes on either side of $\om_0$ to ensure $\overline{\om_s - \om_0} =0$, \ie~to bring about exact cancelation of the third term of (\ref{phis}) when one forms $\bar\phi_s$.  Any deviation of $\overline{(\om_s-\om_0)}$ from zero by as small as one part in $10^{18}$ would be sufficient to cause this third term to become $O(1)$, thereby jeopardizing the effort.  Again as before, since we are working on one contemporaneously acquired data set within the same exposure time $T$, this is not really a problem.  In fact, only one clock will be used to digitize the incident waveform and FFT it, and no relative error among the mode frequencies is actually expected.  The only error is in the absolute frequency of each mode, since this is subject to the intrinsic error of the clock itself, which is irrelevant to the problem\footnote{Although such an error will not affect our effort, we comment in passing that the most accurate clock to date is in any case
stable to 2 parts in 10$^{16} \sqrt{T}$ where $T$ is the integration time in seconds, \cite{sch17}.  Obviously, the exposure time of $T\ap 6$~s we quoted earlier was merely for illustrative purposes, \ie~a more realistic exposure of $T\ap$~6,000~s would increase the stability to one part in $10^{18}$ which would suit the requirement stipulated above, even in terms of absolute error.}.

%Thus it is important that the devices used to constrain the frequency of the modes are all synchronized to the same clock, in which case the relative excursions of the mode frequencies w.r.t. each other could easily be reduced to the necessary level.  Fortunately, the metrological technology available today allows the synchronization of multiple devices to the same master clock with comparable precision. ({\bf Lingze, please comment}).

Ultimately, the goal is to add the {\it same} constant to (or subtract it from) the average phase of each subset of modes to ensure the resultant is proportional to the spectral variance $\si^2$ of the subset with the appropriate corresponding constant of proportionality.  The uncertainty in the dispersive column density $n_e z$ as inferred from (\ref{kpp}) is then determined by the degree to which the aforementioned additive (or subtractive) constant varies from subset to subset.  Thus, there can be no sensible claim of a detection of the ionized intergalactic medium unless the variation in this constant is less than the $k_0^{''}$ term in (\ref{kpp}).

\section{Summary and Conclusion}

The recent method proposed by \cite{lieu13} to detect intergalactic baryons by looking for their dispersive effect on the light of incoherent distant sources relies on a generalization of the treatment of \cite{cor76}, to also include photons as the sort of pulses (or wave packets) susceptible to broadening by dispersion.  In this paper we demonstrate more formally the argument of \cite{hir14}, on how \cite{lieu13} was fundamentally flawed, namely photons emitted by incoherent sources are the consequence of the normal ordering of operators in the second quantization process, \ie~they are not at all the same as the classical pulses of \cite{cor76}. If these classical pulses, each of which is a coherent bundle of many photons, are completely absent from the incoherent source, then the observation strategy of \cite{lieu13} that focusses upon the distortion of intensity fluctuations as a sign of dispersion by the ionized intergalactic baryons will fail.

This paper did not end in such an unfortunate note, however.  We proceeded to suggest another completely different approach, by recognizing that current technology allows one to measure the phase of individual modes of incoherent radiation to test for the presence of the quadratic spectral phase left behind by dispersion as imprint.
%For the time being our attention is confined to Galactic radio sources as means of probing the baryonic content of the interstellar medium, as cosmological sources involve expansion and redshift, which complicates the treatment and can only be investigated in a later paper.  Right now, however, we are able to
We provided observers with a clear strategy that can lead to a robust detection of the quadratic phase, and hence affords them a means of clinching the line-of-sight column density of ionized intergalactic plasma in virtually any direction that ends at a radio source.

\end{document}